\documentclass[conference]{IEEEtran}
\IEEEoverridecommandlockouts
\usepackage{cite}
\usepackage{amsmath,amssymb,amsfonts}
\usepackage{graphicx}
\usepackage{textcomp}
\usepackage{xcolor}
\usepackage{url}
\usepackage{hyperref}
\usepackage{xspace}

\def\BibTeX{{\rm B\kern-.05em{\sc i\kern-.025em b}\kern-.08em
    T\kern-.1667em\lower.7ex\hbox{E}\kern-.125emX}}
    
\usepackage[toc,acronym,nowarn]{glossaries}
\newacronym{OS}{OS}{Operating System}
\newacronym{SE}{SE}{Secure Element}
\newacronym{TCB}{TCB}{Trusted Computing Base}
\newacronym{CA}{CA}{Certificate Authority}
\newacronym{PKI}{PKI}{Public Key Infrastructure}
\newacronym{MITM}{MITM}{Man-In-The-Middle}
\newacronym{TLS}{TLS}{Transport Layer Security}
\newacronym{VMA}{VMA}{Virtual Memory Area}
\newacronym{TPM}{TPM}{Trusted Platform Module}
\newacronym{KVM}{KVM}{Kernel-based Virtual Machine}
\newacronym{VM}{VM}{Virtual Machine}
\newacronym{HV}{HV}{Hypervisor}
\newacronym{TLB}{TLB}{Translation Lookaside Buffer}
\newacronym{PoC}{PoC}{Proof of Concept}
\newacronym{VMCB}{VMCB}{Virtual Machine Control Block}
\newacronym{SLAT}{SLAT}{Second Level Address Translation}
\newacronym{SSH}{SSH}{Secure Shell}
\newacronym{RSA}{RSA}{Rivest–Shamir–Adleman}
\newacronym{AES}{AES}{Advanced Encryption Standard}
\newacronym{OOM}{OOM}{Out Of Memory}
\newacronym{MKTME}{MKTME}{Multi-Key Total Memory Encryption}
\newacronym{VMI}{VMI}{Virtual Machine Introspection}
\newacronym{HSM}{HSM}{Hardware Security Module}
\newacronym{TEE}{TEE}{Trusted Execution Environment}
\newacronym{SGX}{SGX}{Software Guard Extensions}
\newacronym{TA}{TA}{Target Application}
\newacronym{TZ}{TZ}{TrustZone}
\newacronym{HA}{HA}{Host Application}
\newacronym{BIOS}{BIOS}{Basic Input Output System}
\newacronym{ROP}{ROP}{Return Oriented Programming}
\newacronym{TSX}{TSX}{Transactional Synchronization Extensions}

\makeglossaries    
    
\begin{document}

\title{Scanclave: Verifying Application Runtime Integrity in Untrusted Environments\\
}

\author{\IEEEauthorblockN{Mathias Morbitzer}
\IEEEauthorblockA{\textit{Fraunhofer AISEC} \\
Garching near Munich, Germany\\
mathias.morbitzer@aisec.fraunhofer.de}
}

\maketitle

\graphicspath{{figures/}}

\newcommand{\name}{Scanclave\xspace}

\begin{abstract}
  Data hosted in a cloud environment can be subject to attacks from a higher privileged adversary, such as a malicious or compromised cloud provider. 
To provide confidentiality and integrity even in the presence of such an adversary, a number of \glspl{TEE} have been developed. 
A \gls{TEE} aims to protect data and code within its environment against high privileged adversaries, such as a malicious operating system or hypervisor.

While mechanisms exist to attest a \gls{TEE}'s integrity at load time \cite{anati2013innovative}, there are no mechanisms to attest its integrity at runtime. 
Additionally, work also exists that discusses mechanisms to verify the runtime integrity of programs and system components. 
However, those verification mechanisms are themselves not protected against attacks from a high privileged adversary.
It is therefore desirable to combine the protection mechanisms of \glspl{TEE} with the ability of application runtime integrity verification. 

In this paper, we present \name, a lightweight design which achieves three design goals:
Trustworthiness of the verifier, a minimal trusted software stack and the possibility to access an application's memory from a \gls{TEE}.
Having achieved our goals, we are able to verify the runtime integrity of applications even in the presence of a high privileged adversary.

We refrain from discussing which properties define the runtime integrity of an application, as different applications will require different verification methods.  
Instead, we show how \name enables a remote verifier to determine the runtime integrity of an application. 
Afterwards, we perform a security analysis for the different steps of our design. 
Additionally, we discuss different enclave implementations that might be used for the implementation of \name.

\end{abstract}


\glsresetall

\section{Introduction}
\label{sec:introduction}

Cloud computing has been on the rise over the last years. 
This technology allows organizations to outsource their infrastructure and its management to a third party, and to focus on other tasks. 

However, by outsourcing the infrastructure, cloud customers are required to trust the cloud provider not to violate the confidentiality or integrity of their data. 
A malicious provider is trivially able to obtain critical information from its customers, or even to manipulate data and processes. 
This is one of the reasons for the deployment of multiple \glspl{TEE}. 

A \gls{TEE} is an isolated environment which aims to protect executions within its environment against high privileged adversaries. 
While software \glspl{TEE} solely rely on software mechanisms for protection, hardware \glspl{TEE} make use of additional hardware mechanisms to protect the confidentiality and integrity of code and data within the environment. 
The approach of hardware \glspl{TEE} minimizes the entities the cloud customer has to trust to a single one: the hardware vendor.

Most \glspl{TEE} provide the possibility to attest the integrity of their components at load time \cite{anati2013innovative}.
Before the \gls{TEE} is loaded, all required components can be measured, for example by hashing them. 
Those measurements can be compared with previous measurements, which are known to reflect a trustworthy state of the components.
If all hashes are equal to previous measurements, it can be concluded that the \gls{TEE} was loaded by using only trustworthy components, and was therefore in a trustworthy state at load time. 

However, \glspl{TEE} lack mechanisms to verify the runtime integrity of the environment. 
Once the \gls{TEE} - or any other application - is running, its memory is subject to change, for example by allocating memory, or by changing variables. 
At runtime, the state of the application is therefore different from its state at load time. 
Runtime integrity verification is required to determine if this new state is still trustworthy, or if it has been maliciously modified.  

Work exists that discusses how the runtime integrity of applications \cite{haldar2004semantic, davi2009dynamic} and system components \cite{kil2009remote} can be verified. 
Unfortunately, those evaluation mechanisms themselves are not protected against a higher privileged adversary. 
We eliminate this threat by running the evaluation mechanisms in a \gls{TEE}.

In this work, we present \name, an architecture which allows us to achieve the following goals:

\begin{enumerate}
  \item{
    \textbf{Trustworthiness of the verifier}\\
    \name needs to be trustworthy.
    This trustworthiness can not be affected by an adversary, even if he has high privileges on the system.
  }
  \item{
    \textbf{Minimal trusted software stack}\\
    The code that needs to be trusted should be reduced to a minimum.
    This will reduce the attack surface and simplify examination of the code. 
  }
  \item{
    \textbf{Accessing an application's memory from a \gls{TEE}}\\
    \name needs to be able to access an application's memory at runtime.
    This is necessary to perform the runtime integrity measurements. 
  }  

\end{enumerate}

\section{Attacker Model}
\label{sec:attacker}

In this section, we introduce our attacker model and its capabilities. 
The goal of the attacker is to modify an application at runtime without being detected.  

We consider an attacker who is able to gain control over an application after it has been correctly loaded, for example by making use of a vulnerability. 
Having gained control over the application, he is able to perform arbitrary modifications to it, such as modifying return statements or changing code.

We also consider an attacker who is in control of the high privileged software hosting the \gls{TEE}, such as the \gls{OS}. 
The attacker is able to observe the \gls{TEE} with any methods available to the \gls{OS}, for example by influencing its scheduling or by monitoring its network traffic. 
He can also read and modify all memory regions and processes that are not protected by the \gls{TEE}. 
However, the attacker is not able to break any cryptographic primitives or to perform any zero-day attacks on the \gls{TEE}. 
The \gls{TEE} is therefore the only component on the system that can be trusted. 

\section{Design}
\label{sec:design}

\name allows a remote party to verify the runtime integrity of applications, even in the presence of a high privileged adversary.
In this section, we start with explaining how we achieve our goals. 
Afterwards, we describe the design of \name and discuss which steps have to be taken by a remote verifier to perform runtime integrity verification of an application. 

To achieve our first goal, trustworthiness, we embed \name in a \gls{TEE}. 
This will protect \name against higher privileged adversaries. 

To achieve our second goal, a minimal trusted software stack, we use enclaves as \gls{TEE} for the implementation of \name. 
An example for such an enclave is Intel \gls{SGX} \cite{costan2016intel}.
The goal of enclaves is to allow an application in user space to create an area which is protected against software running on higher privilege levels. 
This protected area is called the enclave. 

The enclave application is split up in the untrusted \gls{HA} and the trusted enclave. 
The \gls{HA} can call functions in the enclave by using specified entry points. 
Also the enclave can call functions in the \gls{HA}, for example to perform network communication.
Other than using the specified entry points, the \gls{HA} and high privileged software have no possibility to access or modify an enclave's memory. 
This will protect the enclave even against a high privileged adversary. 

The trusted software stack of enclaves only consists of the code running directly in one particular enclave. 
We move only the code required for verification and attestation into the enclave. 
All operations that can be performed by untrusted code, such as network communication, are placed outside of the \gls{TEE}. 
Those operations have to make use of protection mechanisms such as encryption to ensure their confidentiality and integrity.

While the \gls{HA} has no possibility to access the enclave's memory, the enclave is able to access the memory of its \gls{HA}. 
When the application to be verified, the \gls{TA}, is running within the \gls{HA}, the enclave will be able to access the \gls{TA}'s memory, and to verify its runtime integrity. 
This allows us to achieve the goal of accessing an application's memory from a \gls{TEE}.  

Once we enabled \name to verify the integrity of the \gls{TA}, we can apply a number of verification methods. 
Work on verification methods that ensure runtime integrity of applications already exists \cite{haldar2004semantic, davi2009dynamic, abadi2009control}, and will therefore not be discussed in this work. 
Instead, we will focus on the technical difficulties when trying to verify the \gls{TA} from an enclave.

\begin{figure}
  \includegraphics[width=\columnwidth]{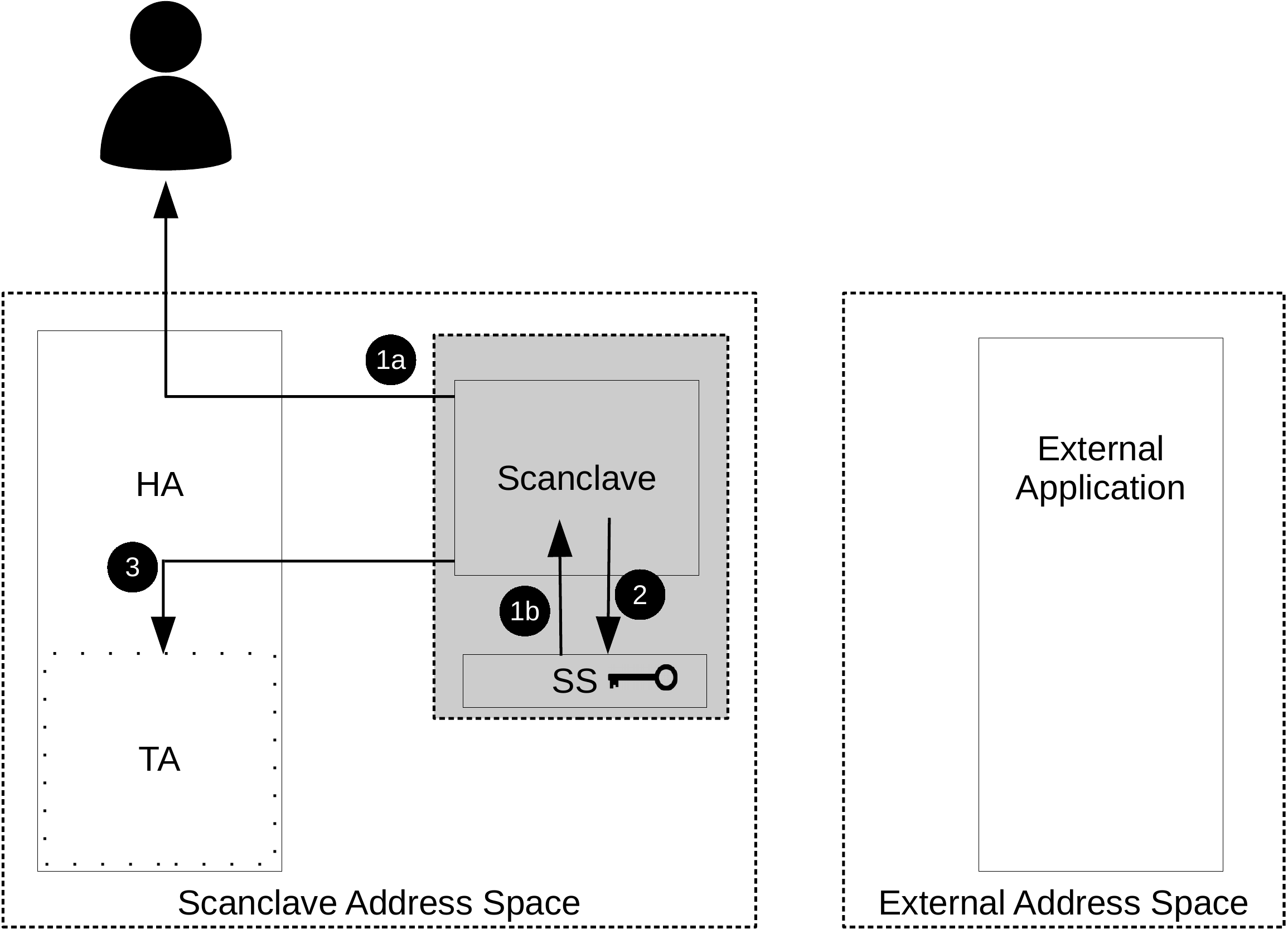}
  \caption{
    The \gls{HA} is used to launch the \gls{TA}. 
    This causes \name and the \gls{TA} to share the same address space, and allows \name to access the \gls{TA}'s memory.  
  }
  \label{fig:initialization}
\end{figure}

Figure \ref{fig:initialization} shows an overview of \name's architecture and its setup. 
The shaded area is protected by the enclave. 
First, the \gls{HA} will hand over control to \name for initialization.  
In this step, the \name instance creates a unique private key, which will be used to sign verification reports.

The private key can not be shipped with \name, as high privileged software is able to inspect all components required to launch the enclave \cite{costan2016intel}. 
While the integrity of the components can be guaranteed by remote attestation protocols, their confidentiality can be violated by a high privileged adversary \cite{anati2013innovative}. 
Secrets like private keys will only be protected within the running enclave, or in a secure storage only accessible to the enclave. 

Once the key has been created, its matching public key will be sent to the remote verifier (1a).  
As noted previously, \name does not include functionality to perform network communication.
It therefore has to rely on the \gls{HA} and the underlying \gls{OS} for communication with the remote verifier. 
This communication might be monitored and modified by an attacker. 
Therefore, communication between \name and the remote verifier needs to be encrypted and its integrity has to be protected. 
It also must be ensured that the secure connection is only terminated within the enclave, and not in the \gls{HA}.

The public key of the remote verifier can be shipped with \name to help establishing a secure communication channel.
Compared to the private key, the public key can be shipped with \name, as only its integrity needs to be ensured.  
Knauth et al. \cite{knauth2018integrating} show how attestation of the enclave can be combined with the establishment of a secure channel. 

After having created the private key and sending the public key to the remote verifier, \name will save the private key in a secure storage only accessible to the particular \name instance (2). 
At future launches, the private key can be retrieved from the secure storage (1b). 
Using a secure storage ensures that the key will never leave the protected area. 
We leave it up to the enclave implementation how the secure storage is implemented, and only require guarantees that only the enclave can access the storage. 
If the enclave implementation does not provide a secure storage, a new private key can be created each time \name is launched.

After the initialization phase, \name will hand control back to the \gls{HA}, which will launch the \gls{TA} (3). 
This step does not require any modification of the \gls{TA}. 
As the \gls{TA} is launched by the \gls{HA}, they will share the same address space together with \name. 
Still, the security mechanisms of the enclave ensure that the memory of \name can not be accessed from the untrusted \gls{TA} or \gls{HA}. 

\begin{figure}
  \includegraphics[width=\columnwidth]{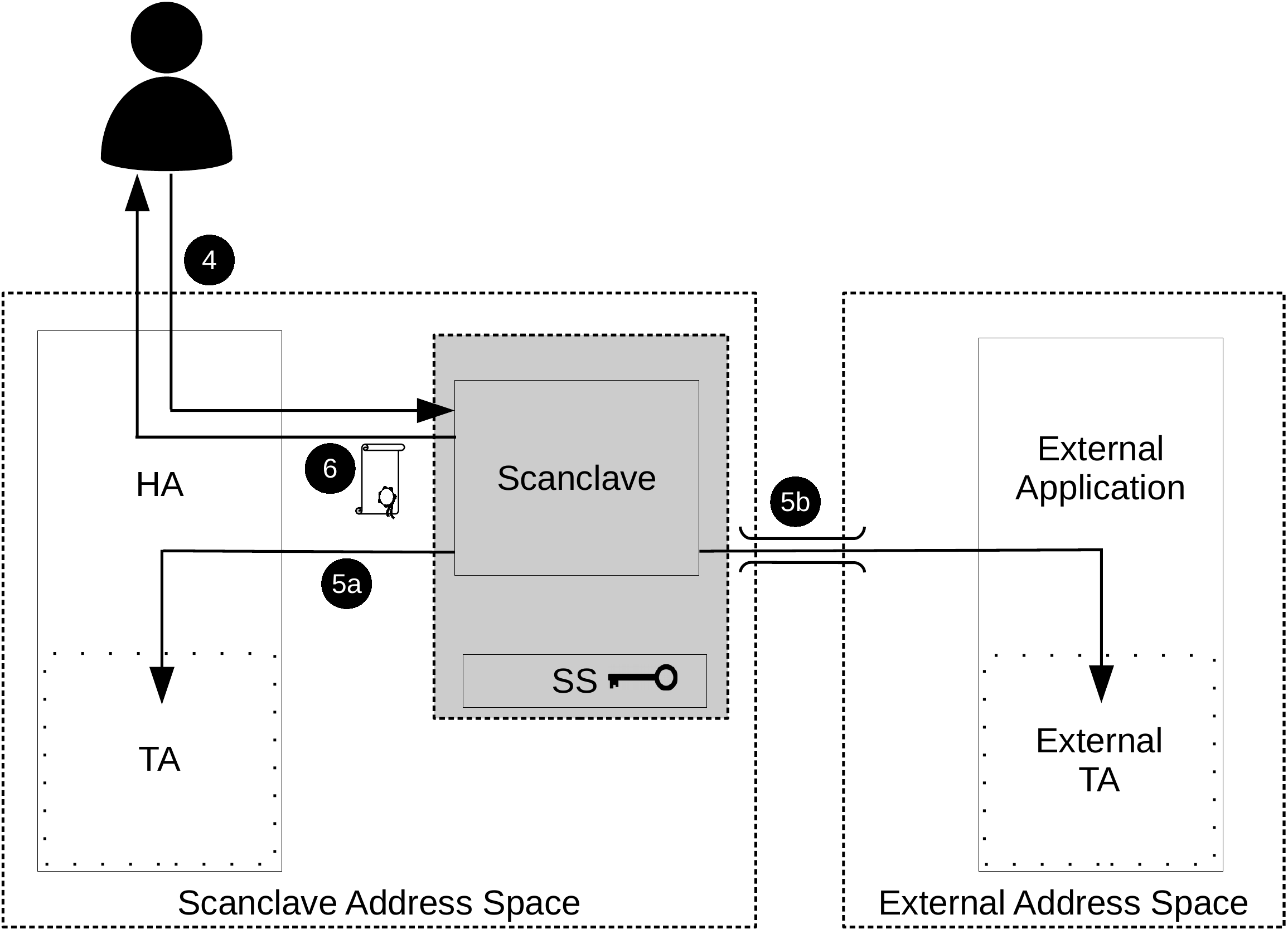}
  \caption{
    A remote verifier can now request \name to verify the runtime integrity of the \gls{TA}. 
    This can be either done by accessing the \gls{TA} directly, or by using a debug bridge.
    The results of the verification will be signed and sent to the verifier.    
  }
  \label{fig:verification}
\end{figure}

Figure \ref{fig:verification} shows the verification process with \name. 
Whenever verification of the \gls{TA} is required, the remote verifier sends a nonce to \name (4). 
\name will then scan the \gls{TA}'s address space and perform all checks required to verify the integrity of the \gls{TA} (5a).

On systems which allow applications to access the memory of other applications, \name might also be used to verify an external \gls{TA}. 
To access an external \gls{TA}'s memory, \name can attach itself to the external application by creating a debug bridge (5b). 
Such a bridge can be created for example by using \texttt{ptrace} \cite{haardt1993ptrace} on Linux or the \texttt{ReadProcessMemory} \cite{microsoft2019read} function on Windows. 

We use \name only to verify the \gls{TA}, but not the \gls{HA}.
Verification of the \gls{HA} is not required, as it is only used to host \name and to launch the \gls{TA}. 

After gathering the verification results and creating a verification report, the report is signed with the unique private key of the \name instance. 
The signed report will be sent to the remote verifier (6). 
This allows the remote verifier in possession of the public key to verify that the report was created by the correct \name instance.
To defend against replay attacks, the nonce sent in Step (4) is included in the report. 
Steps (4) to (6) can be repeated whenever verification is desired.  

If the verification report shows that the \gls{TA} is not in a trustworthy state, the remote verifier can restore the trustworthy state of the \gls{TA} with the help of \name.  
One approach would be to try to determine all modifications to the \gls{TA}, and to restore its original state. 
However, it can be challenging to restore all malicious modifications to the \gls{TA}. 
Instead, \name instructs the \gls{HA} to terminate the \gls{TA}, and to launch it again in a clean state. 

An alternative approach to our design would be to run the whole \gls{TA} in the enclave. 
In this scenario, an attacker compromising the \gls{TA} would imply a compromise of the enclave.  
Software running in the enclave could therefore not be trusted anymore. 
This violation would be impossible to detect with standard attestation techniques for \glspl{TEE}, as they only verify load time integrity. 
By using \name, an attacker gaining control over the \gls{TA} can be detected by runtime integrity verification. 
The confidentiality and the integrity of \name are protected against a higher privileged attacker, and the confidentiality of the \gls{TA} is assured by \name. 

\section{Discussion}
\label{sec:discussion}

This section discusses how an adversary might try to influence the different steps of  \name's verification process.

When transferring the public key to the remote verifier after initialization in Step (1a), an adversary might try to impersonate \name, or to exchange \name's public key with its own. 
This would allow the adversary to create arbitrary verification reports in the name of \name. 
It is therefore important to ensure the remote verifier is communicating with a valid \name instance, and to ensure the integrity of the communication. 

To avoid an adversary impersonating a \name instance, the remote verifier uses a remote attestation protocol to ensure he is communicating with a valid \name instance \cite{anati2013innovative}.
This also allows to detect a high privileged adversary that prevents \name from launching. 
To ensure that an adversary is not able to modify the public key sent to the remote verifier, the secure communication channel is only terminated within the enclave \cite{knauth2018integrating}. 
The same mechanisms also ensure the integrity of the nonce sent to \name in Step (4). 
The verification report sent in Step (6) is additionally signed with \name 's private key.

Steps (1b) and (2), reading and writing the private key from and to the secure storage, are performed within the protected area. 
We do rely on the enclave implementation to protect accesses from the enclave to the secure storage against adversaries.

In Step (3), the \gls{HA} launches the \gls{TA}. 
An adversary could try to prevent the launch, or to launch a modified version of the \gls{TA}. 
If the \gls{TA} is not launched, it will be detected by \name as it will not be able to find the \gls{TA} in its address space. 
For external \glspl{TA}, \name will detect that it is not able to build a debug bridge to the \gls{TA}.
If the \gls{TA} has been modified, this will be detected by the verification mechanisms of \name.

Before the verification process in Steps (5a) and (5b), an adversary might try to read or modify the \gls{TA}'s memory, or try to manipulate the runtime verification performed by \name. 
Reading the \gls{TA}'s memory will be possible, as our design does not provide the \gls{TA} with confidentiality. 
In case confidentiality is required for the \gls{TA}, a second enclave for handling confidential data can be hosted. 
Unlike reading the \gls{TA}'s memory, modification of the \gls{TA} will be detected by the runtime verification. 
An attacker might therefore try to avoid detection through the verification process. 
For doing so, an attacker has two possibilities: 

\textbf{Causing undetectable changes.}
Our approach provides a general design for performing runtime verification. 
An attacker trying to avoid being detected by the verification process could try to only apply modifications to the \gls{TA} that can not be detected. 
It is therefore important to choose a verification method that is able to detect different kinds of modifications, or to combine multiple methods. 
We leave it up to future work to analyze and compare existing verification methods. 

\textbf{Restoring integrity before verification.}
Another method to avoid detection by \name would be to restore the original state of the \gls{TA} before runtime integrity verification is performed. 
To find out when the verification is performed, the attacker could analyze the enclave's behavior, such as its memory access patterns or network traffic. 
Having determined the moment of the verification, the attacker could momentarily restore the \gls{TA}'s integrity to ensure a successful verification. 

It is also important to keep in mind that accesses from the enclave to the \gls{TA}'s memory are not protected as they are performed on untrusted memory regions.
An attacker could try to redirect \name's verification memory accesses to other memory regions, in which an unmodified copy of the \gls{TA} is stored. 

To avoid an adversary finding out the moment of the verification, it is important not to leak any information of the enclave's behavior, such as memory or cache access patterns. 
Previous work covers how software can protect itself against such leakage \cite{stefanov2013path, shih2017tsgx, seo2017sgx, awad2017obfusmem, dong2018shielding}. 

An adversary might also monitor the network traffic and wait for the remote verifier to send the nonce in Step (4). 
He could then conclude that Step (5), the verification, will start shortly. 
To avoid that the timing of Step (5) can be determined by observing Step (4), \name performs the verification independently of Step (4). 
Instead, Step (4) is performed within a regular interval, which is regularly changed to avoid predictability. 
If verification has been performed and a new nonce was received, the current verification report is sent to the remote verifier, using the new nonce.  

To sum up, all steps in our process are protected against an adversary monitoring the network traffic as well as against adversaries trying to modify the \gls{HA}, \gls{TA}, or \name.

\section{Implementation considerations}
\label{sec:implementation}

In this section, we discuss the state of the art in enclave implementations in the industry and in academia.
We analyze their suitability for our approach as well as their limitations.

Among all enclave \glspl{TEE}, Intel \gls{SGX} is probably the most widely known. 
One reason for this is that it was also one of the first \glspl{TEE}, proposed in 2015. 
Additionally, the fact that it is the only enclave \gls{TEE} available on the x86 architecture helps the publicity of \gls{SGX}. 
Meanwhile, even the first cloud providers allow customers to use \gls{SGX} \cite{azure2017sgx, ibm2017sgx}.
Still, the spread of \gls{SGX} is limited by various factors. 
One of them is that the underlying BIOS needs to support \gls{SGX} \cite{intel2017detecting}. 
Until recently, to make use of an \gls{SGX} implementation with all available security guarantees, a business relationship with Intel was required. 
This step was necessary to make use of the attestation process required to launch enclaves \cite{anati2013innovative}. 
After this requirement has received criticism in the past \cite{costan2016intel}, Intel recently announced to allow attestation also for third parties \cite{scarlate2018supporting}. 

Various attacks on \gls{SGX} have been made public. 
While some analyze the memory access pattern of the enclave to extract secrets \cite{xu2015controlled, van2017telling}, others are able to reconstruct secrets via a caching side-channel \cite{brasser2017software,gotzfried2017cache,van2018foreshadow}. 

To develop an enclave that is protected against side-channel attacks, Costan et al. propose Sanctum \cite{costan2016sanctum}. 
The basic architectures of \gls{SGX} and Sanctum are very comparable. 
By isolating an enclave's cache within the system's cache hierarchy, Sanctum additionally protects against the caching side-channel.
It also protects against attacks that analyze an enclave's memory access pattern by deploying a page-coloring-based cache partitioning scheme.

Sanctum was designed as an extension to the RISC-V architecture and is open source. 
Open sourcing the security mechanisms allows to investigate the mechanisms which ensure the trustworthiness of the enclave. 
The main limitation of Sanctum is that it is making use of a non-standard hardware extension. 

For this reason, the Keystone enclave was introduced in 2018. 
Similar to Sanctum, Keystone is also built on the RISC-V architecture and open source. 
Additionally, one of its main goals is to enforce memory isolation only by using standard RISC-V primitives.
This reduces the barriers when adopting Keystone. 
Keystone also has the advantage that it can be run in a \gls{VM}, which allows to evaluate and develop enclaves without needing specific hardware. 

\begin{table}[htbp]
  \caption{Comparison of \gls{SGX}, Sanctum and Keystone}
  \begin{center}
  \begin{tabular}{| p{4cm} | c | c | c |}
     \hline
     & \gls{SGX} &  Sanctum & Keystone\\
     \hline
     Architecture & x86 & RISC-V & RISC-V\\
     Open source & No & Yes & Yes\\
     Attestation infrastructure required & Yes & No & No\\
     Available in cloud & Yes & No & No\\
     Memory access pattern protection & No & Yes & In Progress\\
     Caching side-channels protection & No & Yes & In Progress\\
     Hardware extensions required & No & Yes & No\\
     \hline
  \end{tabular}
  \label{tab:enclaves}
  \end{center}
\end{table}

In Table \ref{tab:enclaves}, we give an overview of the differences between \gls{SGX}, Sanctum and Keystone. 
Considering the limitations of \gls{SGX}, and the hardware requirements of Sanctum, we expect that an implementation of \name on Keystone would be the most promising. 

\section{Related Work}
\label{sec:related}

To the best of our knowledge, we are not aware of any work that covers application runtime integrity verification from enclaves. 
We will therefore start with discussing work that also makes use of the possibility to access the \gls{HA}'s memory from the enclave. 
We will continue with work that was designed to verify system integrity with the help of \glspl{TEE}.

Schwarz et al. \cite{schwarz2019practical} presented malware running in an \gls{SGX} enclave. 
As an enclave is able to access the memory of its \gls{HA}, they built a malicious enclave which impersonates its \gls{HA}.
First, they use a technique based on Intel TSX\cite{intel2015tsx} to look for code fragments in the \gls{HA}'s memory that can be used to perform \gls{ROP} attacks \cite{shacham2007geometry}. 
Having discovered interesting code fragments, they use a technique dubbed SGX-ROP to execute arbitrary code in the \gls{HA}. 

Their attacker model greatly differs from ours, as they assume a malicious enclave, and a benign \gls{HA} as well as benign high privileged software. 
In contrast, \name is designed to protect a benign \gls{TA} against malicious high privileged software with the help of a trusted enclave. 
Still, their work is relevant for the implementation of verification mechanisms in \name as it for example discusses techniques to scan a \gls{HA}'s address space from an enclave. 

ARM \gls{TZ} \cite{arm2019trustzone} splits the whole system in a secure and a normal world. 
While the normal world is not able to access the secure world, the secure world is given full control over the normal world. 
Xinyang and Jaeger \cite{xinyang2014sprobes}, and Azab et al. \cite{azab2014hypervision} make use of this property and add methods to the secure world to protect the integrity of a kernel running in the normal world. 

\gls{TZ} was designed with a focus on embedded systems.
It is difficult to apply it in our scenario as it only provides one secure world on each system. 
Different users would all have to run their verification mechanisms in the same secure world. 
This would require to trust the code of other users. 
With enclaves, each \name instance is provided with its own \gls{TEE}, isolated from other \name instances. 
Additionally, code in the secure world receives excessive privileges. 
A cloud provider will want to avoid granting those privileges to a customer. 
With enclaves, the user's verification software is running with lower privileges, protecting the system against a possible malicious cloud user. 

Zhang et al. \cite{zhang2002secure} proposed to make use of a secure coprocessor for integrity verification. 
Such a coprocessor is running independently from the processor on the system, and is therefore unaffected by a compromise of the system.
Petroni et al. \cite{petroni2004copilot} implemented a system which makes use of a secure coprocessor, and used it to monitor the integrity of the host kernel.

Those mechanisms were designed for a different use case, namely for administrators to verify their systems. 
This differs from our approach, which is designed for users wanting to protect their applications against a high privileged adversary. 
Therefore, they have the same limitations in our use case as \gls{TZ}: only one \gls{TEE} exists, and the software in the \gls{TEE} receives extensive privileges. 

The \gls{TPM} \cite{tcg2011tpm} is an industry standard for a micro controller designed to help detecting modifications of a system. 
By creating a hash of every component in the boot chain and storing the hash in the \gls{TPM}, changes to the boot chain can be detected \cite{microsoft2012secured}. 
However, this approach suffers from the problem that the hashes change each time a component in the boot chain is updated. 
Secure boot \cite{arbaugh1996secure} tries to avoid this problem by verifying the signature of components instead of their hashes.

Both methods are designed to verify the boot process rather than applications. 
Infrastructure also exists to extend the boot chain to the application layer \cite{sailer2004ima}.
However, this infrastructure only measures the load time integrity of applications, and requires a benign \gls{OS}, which we do not assume in our design. 

The \gls{TPM} can also be used to establish a dynamic root of trust, which allows to perform attestation after the system was booted. 
To perform such an attestation, Flickr \cite{mccune2007execution} makes use of Intel TXT \cite{greene2012txt} or AMD SVM \cite{amd2005svm} technology. 
Flickr calculates hashes of critical system components and stores them in the \gls{TPM}, where they can be used for attestation with an external entity. 

This approach requires cooperation of high privileged software to execute the CPU instructions required to perform the dynamic attestation. 
During the attestation phase, actions that could interfere with this process, such as debugging and interrupts, are disabled. 
This might affect other computations, which is critical in cloud environments.  
Additionally, most systems are equipped with only one \gls{TPM}. 
To store hashes for different users, multiple \glspl{TPM} or virtualization would be required. 
For those reasons, TXT in cloud environments is mostly used to verify the integrity of system components such as the BIOS or the \gls{HV} \cite{intel2013txt}. 
Our design enables a user to perform application verification without requiring high privileged software, and without affecting other computations on the system. 

\section{Conclusion and future work}
\label{sec:conclusion}

In this work, we presented the design of \name. 
Our design achieves three main goals: Trustworthiness of the verifier, a minimal trusted software stack and the ability to access an application's memory from a \gls{TEE}.
Achieving all three goals, we are able to provide integrity in environments in which higher privileged software can not be trusted, such as cloud environments.

Based on our design, we showed which steps are necessary for a remote verifier to perform application runtime integrity verification. 
Additionally, we discussed how we protect \name against different adversaries in every step of the process. 
Our contribution is completed with a comparison of enclave implementations that could be used for the implementation of \name. 

In future work, we plan to use the \name architecture to develop new and effective verification techniques that can determine the runtime integrity of various types of applications.
One possible scenario would be to use a \gls{VM} as \gls{TA}. 
\name would then be able to measure the integrity of the \gls{VM}'s kernel, as for example as described in \cite{kil2009remote}.

\bibliographystyle{splncs04}
\bibliography{biblio}

\end{document}